\documentclass[english,aps,prl,twocolumn,groupedaddress,superscriptaddress,nofootinbib]{revtex4}
\usepackage[T1]{fontenc}
\usepackage[latin9]{inputenc}
\usepackage{amsmath}
\usepackage{amssymb}
\usepackage{amsfonts}
\usepackage{graphicx}
\usepackage{hyperref}
\usepackage{color}
\usepackage{mathrsfs}
\usepackage{multirow}
\usepackage{float}
\usepackage{bm}
\usepackage{dsfont}
\usepackage{txfonts}

\makeatletter

\definecolor{nblue}{rgb}{0.3,0.3,1.0}
\definecolor{ngreen}{rgb}{0.2,0.7,0.2}
\definecolor{nred}{rgb}{0.9,0.1,0}
\definecolor{nblack}{rgb}{0,0,0}

\newcommand{\beq}{\begin{equation}}
\newcommand{\eeq}{\end{equation}}
\newcommand{\bqa}{\begin{eqnarray}}
\newcommand{\eqa}{\end{eqnarray}}

\makeatletter

\usepackage{babel}
\begin{document}

\title{Multipartite Einstein-Podolsky-Rosen steering sharing with separable states}
\author{Yu~Xiang}
\address{State Key Laboratory of Mesoscopic Physics, School of Physics, Peking University, Collaborative Innovation Center of Quantum Matter, Beijing 100871, China}
\address{Collaborative Innovation Center of Extreme Optics, Shanxi University, Taiyuan 030006, China}
\author{Xiaolong Su}
\address{Collaborative Innovation Center of Extreme Optics, Shanxi University, Taiyuan 030006, China}
\address{State Key Laboratory of Quantum Optics and Quantum Optics Devices, Institute of Opto-Electronics,
Shanxi University, Taiyuan 030006, China}
\author{Ladislav Mi\v{s}ta, Jr.}
\affiliation{Department of Optics, Palack\'y University, 17. listopadu 12, 771 46 Olomouc, Czech Republic}
\author{Gerardo~Adesso}
\address{Centre for the Mathematics and Theoretical Physics of Quantum Non-Equilibrium Systems (CQNE), School of Mathematical Sciences, The University of Nottingham, Nottingham NG7 2RD, United Kingdom}
\author{Qiongyi~He}
\address{State Key Laboratory of Mesoscopic Physics, School of Physics, Peking University, Collaborative Innovation Center of Quantum Matter, Beijing 100871, China}
\address{Collaborative Innovation Center of Extreme Optics, Shanxi University, Taiyuan 030006, China}

\begin{abstract}
 Distribution of quantum correlations among remote users is a key procedure underlying many quantum information technologies. Einstein-Podolsky-Rosen steering, which is one kind of such correlations stronger than entanglement, has been identified as a resource for secure quantum networks. We show that this resource can be established between two and even more distant parties by transmission of a system being separable from all the parties. For the case with two parties, we design a protocol allowing to distribute one-way Gaussian steering between them via a separable carrier; the obtained steering can be used subsequently for one-sided device-independent (1sDI) quantum key distribution. Further, we extend the protocol to three parties, a scenario which exhibits richer steerability properties including one-to-multimode steering and collective steering, and which can be used for 1sDI quantum secret sharing. All the proposed steering distribution protocols can be implemented with squeezed states, beam splitters and displacements, and thus they can be readily realized experimentally. Our findings reveal that not only entanglement but even steering can be distributed via communication of a separable system. Viewed from a different perspective, the present protocols also demonstrate that one can switch multipartite states between different steerability classes by operations on parts of the states.
\end{abstract}
\maketitle

Einstein-Podolsky-Rosen (EPR) steering was put forward by Schr\"{o}dinger~\cite{Schrodinger35} to describe the ``spooky action-at-a-distance'' phenomenon discussed in the original 1935 EPR paradox~\cite{EPR35,Reid89}, which allows one observer to remotely adjust (`steer') the state of another distant observer by local measurements. This special type of quantum correlations offers insights into directional nonlocality \cite{ReidRMP,one-way-Theory, He15,
Adesso15,ReidJOSAB,OneWayNatPhot,OneWayPryde,OneWayGuo,OneWayXu,prlSu,prxresource} and differs conceptually from inseparable correlations aka entanglement~\cite{Howard07PRL,Eric09}. The fact that steering enables verification of shared entanglement even when one party's measurements are untrusted ~\cite{Howard07PRA,Eric13,cavalcanti17review} makes it an essential resource for a number of applications, such as one-sided device-independent (1sDI) quantum key distribution (QKD)~\cite%
{1sDIQKD,1sDIQKD_howard,HowardOptica,CV-QKDexp} and quantum secret sharing (QSS)~\cite{ANUexp,YuQSS,GiannisQSS}, secure quantum teleportation~\cite{SQT13Reid,SQT15,SQT16_LiCM}, and subchannel discrimination \cite{subchannel,subchannel16}.

However, in general it is harder to establish EPR steering than entanglement, as the former requires stronger interaction between parties and tolerates less noise than the latter \cite{he2013prl,He15}. In a quantum network, it is not practical to require that all users are capable of producing steerable states, moreover directly distributing steering is unsafe since an eavesdropper may attack the transmitted quantum states and obtain confidential information. A simpler and more efficient scenario would be to have a quantum \textit{Cloud Server} which can generate quantum states and perform appropriate operations for different quantum tasks, and then establish desired correlations between the users mediated by transmission of ancillary systems with minimal resources. Somehow counterintuitively, it has been recently shown theoretically \cite{qubitthe,Gaussianthe,impro_pra} and experimentally \cite{qubitexp,Gaussianexp1,Gaussianexp2} that entanglement can be distributed between two parties via a separable ancilla. As EPR steering is strictly stronger than entanglement, a question that naturally arises is as to whether also steering can be distributed between two or more distant users via a separable system?

\begin{figure}[tb]
\begin{center}
\includegraphics[width=0.8\columnwidth]{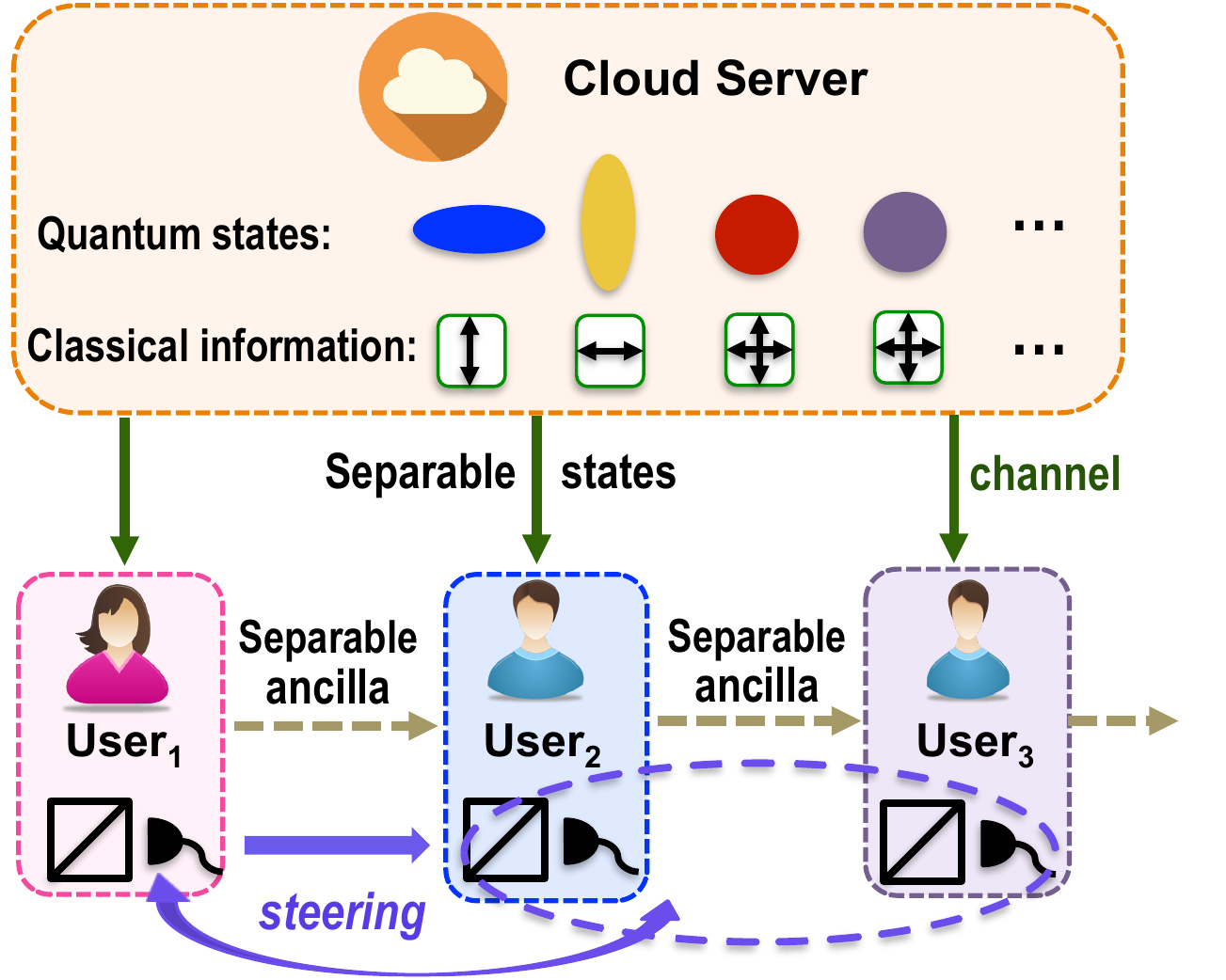}
\protect\caption{Scheme to distribute Gaussian EPR steering among different users via a separable ancilla: The quantum Cloud Server locally produces the quantum states and analyzes the classical information of the displacements required by the task, then sends the separable quantum states to the users. By transmitting a separable ancilla, the users can successfully share EPR steering with desired properties.\label{system}}
\end{center}
\end{figure}

In this Letter, we answer the question in the affirmative. We propose a protocol that can distribute EPR steering among many parties in continuous variable (CV) Gaussian states by transmitting a separable ancilla mode between the users. By preparing locally initial quantum states and performing suitably tailored local correlated displacements on them, as illustrated in Fig.~\ref{system}, we can distribute rich steering properties, such as one-way Gaussian steering in two-users scenario, two-way steering and collective steering which can be used for CV QSS~\cite{QSS,WeedQSS} in three-users scenario, and so on. In particular, we derive analytical thresholds on the displacements as a function of the squeezing degree of the initial states such that the protocol succeeds, and prove that the largest steerability that can be distributed recovers that of the multimode states created by the same optical network without the correlated displacements. This means that the displacements only make the transmitted ancilla mode separable from the rest without reducing the amount of steering. Notably, all the modes used for the distribution are separable from the users, so the eavesdropper cannot decipher any useful information from the channel, making the protocol robust against loss and leakage in long distance transmissions. We further present a modified scheme with a relaxed condition that the ancilla is nonsteerable instead of separable from the users, yielding a broader range of parameters for which the protocol works.

{\it Gaussian steerability.---} For any $(n_A+m_B)$-mode Gaussian state, we put the amplitude (position) and phase (momentum) quadratures of each mode into a column vector $\hat{\xi}:=(\hat{x}_{1}^{A},\hat{p}_{1}^{A},...,\hat{x}_{n_{A}}^{A},\hat{p}_{n_{A}}^{A},\hat{x}_{1}^{B},\hat{p}_{1}^{B},...,\hat{x}%
_{m_{B}}^{B},\hat{p}_{m_{B}}^{B})^{\top}$, satisfying the canonical commutation rules $[\hat{x}_j,\hat{p}_k]=2i\delta_{jk}$. The properties of the state can be fully specified by its covariance matrix (CM) $\gamma_{AB}$ with elements $(\gamma_{AB}) _{ij}=\langle \hat{\xi}_{i}\hat{\xi}_{j}+\hat{\xi}_{j}\hat{\xi}_{i}\rangle /2-\langle \hat{\xi}_{i}\rangle \langle \hat{\xi}%
_{j}\rangle $, which reads as $\gamma_{AB}=\left(
\begin{array}{cc}
A & C \\
C^{\top} & B%
\end{array}
\right).
$
Here, the submatrices $A$ and $B$ are the CMs corresponding to the reduced states of each subsystem, respectively. The steerability from Alice to Bob via Gaussian measurements can be quantified by~\cite{Adesso15}
\begin{equation}  \label{GSAtoB1}
\mathcal{G}^{A\rightarrow B}(\gamma _{AB}):=\max \bigg\{0, \underset{j:\bar{%
\nu}_{j}^{AB\backslash A}<1}{-\sum }\ln (\bar{\nu}_{j}^{AB\backslash A})%
\bigg\},
\end{equation}
where $\bar{\nu}_{j}^{AB\backslash A}$ $(j=1,...,m_B)$ are the symplectic eigenvalues of the Schur complement of $A$ defined as $\bar{\gamma}_{AB\backslash A}=B-C^{\mathsf{T}}A^{-1}C$. The quantity $\mathcal{G}^{A\rightarrow B}$ is a monotone under Gaussian local operations and classical communication \cite{Adesso16} and vanishes when Alice cannot steer Bob by Gaussian measurements \cite{Adesso15,noteGauss}. This quantifier has been experimentally measured in Gaussian cluster states by reconstructing the CM~\cite{prlSu}.

For the sake of experimental feasibility, one can also confirm the presence of steering when the EPR variance product $E_{B|A}:=\Delta_{inf,A}\hat{x}_{B}\Delta_{inf,A}\hat{p}_{B}<1$, where $\Delta_{inf,A}\hat{x}_{B}=\Delta(\hat{x}_{B}-g_x\hat{x}_{A})$ is the minimum inferred variance of Bob's position outcome given the Alice's result with optimal gain factor $g_x=\langle \hat{x}_{B},\hat{x}_{A}\rangle /(\Delta\hat{x}_{A})^2$~\cite{He15}. Here we use the notation $\langle x,y \rangle:=\langle xy\rangle-\langle x\rangle\langle y\rangle$. The $\Delta_{inf,A}\hat{p}_{B}$ is defined similarly. The criterion $E_{B|A}<1$ is necessary and sufficient to test steering by Gaussian measurements. The quantity $E_{B|A}$, which can be easily measured by homodyne detection with high efficiency~\cite{ANUexp}, in fact directly quantifies the Gaussian steerability, as one has $\mathcal{G}^{A\rightarrow B}=\max\{0,\,-\ln E_{B|A}\}$  for all two-mode CMs $\gamma_{AB}$ in standard form (i.e., with diagonal blocks $A,B,C$)~\cite{Adesso15,YuQSS}.

\begin{figure}[tb]
\begin{center}
\includegraphics[width=0.85\columnwidth]{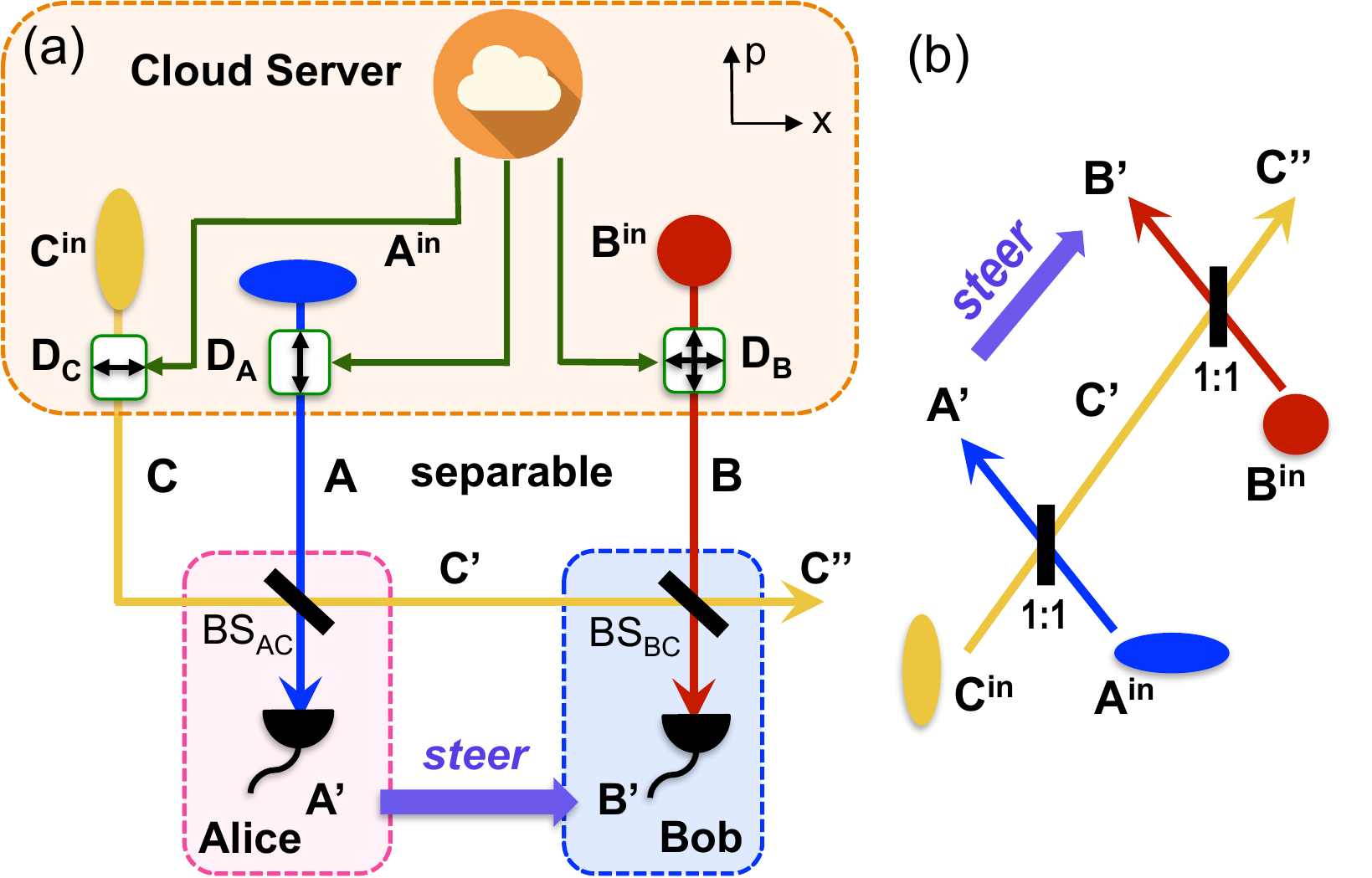}
\protect\caption{(a) Sketch of the one-way Gaussian steering distribution protocol. See text for details. The best steerability is recovering the steerability in the tripartite entangled state created by the optical network, as illustrated in (b).\label{system3mode}}
\end{center}
\end{figure}

In the following, we first show that one-way Gaussian steering by Gaussian measurements can be distributed from Alice to Bob by a separable ancilla. We improve the protocol that was developed to distribute entanglement~\cite{impro_pra} by optimizing the displacements to distribute the highest steerability. As depicted in Fig.~\ref{system3mode}(a), the initial modes $A$, $B$, and $C$ sent from the Cloud Server are in a fully separable state, and the ancillary mode $C'$ is separable from the modes held by Alice and Bob. This ensures the security of the  task, even if the communication channels are  tapped. We prove that the highest steerablity which can be distributed by a separable ancilla is determined by the utilized optical network composed only of input squeezed states and beam splitters, as depicted in Fig.~\ref{system3mode}(b).

{\it Protocol.---} The protocol consists of three steps. In {\it Step 1}, the Cloud Server initially produces modes $A^{\rm in}$ and $C^{\rm in}$ in momentum- and position-squeezed vacuum states, respectively, while mode $B^{\rm in}$ is in a vacuum state. All three modes are in a product state described by the
diagonal CM $\gamma^{\rm in}_{ABC}=\mbox{diag}(e^{2t},e^{-2t},1, 1, e^{-2t}, e^{2t})$, where $t$ is the squeezing parameter. The three modes are then appropriately displaced by local correlated displacements $\hat{p}_{A^{\rm in}}\rightarrow\hat{p}_{A^{\rm in}}-D_Ap_d$, $\hat{x}_{C^{\rm in}}\rightarrow\hat{x}_{C^{\rm in}}+D_Cx_d$, $\hat{x}_{B^{\rm in}}\rightarrow\hat{x}_{B^{\rm in}}+D_Bx_d$, $\hat{p}_{B^{\rm in}}\rightarrow\hat{p}_{B^{\rm in}}+D_Bp_d$. Here, $x_d$ and $p_d$ obey a zero mean Gaussian distribution with the same variance and $D_A,\ D_B,\ D_C$ are the strengths of the displacements, which will be specified in the second step. The resulting state is fully separable with CM $\gamma_{ABC}=\gamma^{\rm in}_{ABC}+\tilde{D}$,
where $\tilde{D}$ denotes a positive noise matrix created by the displacements.

In {\it Step 2}, Alice superimposes modes $A$ and $C$ on a balanced beam splitter $BS_{AC}$ thereby creating a three-mode state with CM $\gamma_{A'BC'} =U_{AC}(\gamma^{\rm in}_{ABC}+\tilde D)U_{AC}^{\top} =U_{AC}\gamma^{\rm in}_{ABC}U_{AC}^{\top}+xP$, $x \geq 0$. Here, the matrix $U_{AC}$ describes the beam splitter. $U_{AC}\gamma^{\rm in}_{ABC}U_{AC}^{\top}$ describes a product state of vacuum mode $B^{\rm in}$ and a two-mode squeezed vacuum state after the transmission of squeezed modes $A^{\rm in}$ and $C^{\rm in}$ free of any displacements through the beam splitter, as depicted in Fig.~\ref{system3mode}(b). However, the entanglement between the output modes $A'$ and $C'$ can be smeared by addition of a sufficiently large nonnegative multiple $xP$, where $P$ is a suitable positive noise matrix and $x$ is an important parameter to adjust the strength of noise. Using the method developed in Ref.~\cite{defineP}, the matrix $P$ can be constructed from the following $6\times1$ vectors $q_1$ and $q_2$ to negate the entanglement between mode $A'$ and $C'$:
\begin{equation}
q_1 =
(0,-1,0,d_B,0,-1)^{\top},\quad
q_2 =
(1,0,d_B,0,-1,0)^{\top}, \label{P}
\end{equation}
as $P=q_1q_1^{\top}+q_2q_2^{\top}$. Note that the variable parameter $d_B$ can be optimized to reach the highest steerability from Alice to Bob in the final state. The CM of the state after {\it Step 2} becomes
\begin{equation}\label{gamma2}
\gamma_{A'BC'}=
\left(\begin{array}{ccc}
m\mathds{1} & d_Bx\sigma_z &n\sigma_z \\
d_Bx\sigma_z & (1+d_B^2x)\mathds{1} & -d_Bx\mathds{1} \\
n\sigma_z & -d_Bx\mathds{1} & m\mathds{1}
\end{array}
\right),
\end{equation}
where $m=\cosh{2t}+x$ and $n=\sinh{2t}-x$. Hence, one can determine the correlation matrix of displacements
$\tilde{D}$ prior to the beam splitter $BS_{AC}$ as $\tilde{D}=xU_{AC}^{\top}PU_{AC}$, which gives
\begin{equation}
\tilde{D}=\left(
\begin{array}{cccccc}
0 & 0 & 0 & 0 & 0 & 0 \\
0 & 2x & 0 & -\sqrt{2}d_Bx & 0 & 0\\
0 & 0 & d_B^2x & 0 &\sqrt{2}d_Bx & 0\\
0 & -\sqrt{2}d_Bx & 0 & d_B^2x & 0 & 0\\
0 & 0 & \sqrt{2}d_Bx &0 & 2x & 0\\
0 & 0 & 0 & 0 & 0 & 0
\end{array}
\right). \label{Q}
\end{equation}
This corresponds to displacement strengths $D_A=D_C=\sqrt{2},\ D_B=d_B$, and displacement variances
$\langle(\Delta x_d)^2\rangle=\langle(\Delta p_d)^2\rangle=x$. The free parameters $d_B$ and $x$
need to be suitably adjusted for the protocol to work, as done in the next step.

In {\it Step 3}, Bob mixes mode $C'$ sent by Alice with his mode $B$ on the balanced beam splitter $BS_{BC}$, which yields the final CM $\gamma_{A'B'C''}=U_{BC}\gamma_{A'BC'}U_{BC}^{\top}$ in the form
\begin{equation}
\gamma_{A'B'C''}=\left(
\begin{array}{ccc}
m\mathds{1} & \frac{d_Bx+n}{\sqrt{2}}\sigma_z & \frac{d_Bx-n}{\sqrt{2}}\sigma_z \\
\frac{d_Bx+n}{\sqrt{2}}\sigma_z & \frac{1+m+d_Bx(d_B-2)}{2}\mathds{1} & \frac{1+d_B^2x-m}{2}\mathds{1} \\
\frac{d_Bx-n}{\sqrt{2}}\sigma_z  & \frac{1+d_B^2x-m}{2}\mathds{1} & \frac{1+m+d_Bx(d_B+2)}{2}\mathds{1}
\end{array}
\right). \label{gamma3}
\end{equation}
Now, by quantifying the amount of the distributed Gaussian steering from mode $A'$ to mode $B'$ via Eq.~(\ref{GSAtoB1}) we get
\begin{equation}\label{GAtoB}
\mathcal{G}^{A'\rightarrow B'}=\ln [2\cosh{2t}/(\cosh{2t}+1)]\,,
\end{equation}
with optimal displacement $d_B^{opt}=\tanh{2t}+1$. Note that $\mathcal{G}^{A'\rightarrow B'}>0$ for any $t>0$. Interestingly, the right-hand side of Eq.~(\ref{GAtoB}) equals the amount of steerability from $A'$ to $B'$ in the scheme without any displacements, shown in Fig.~\ref{system3mode}(b)~\cite{supp}. This means that the optimal displacements ensure separability of the transmitted ancilla from the other modes, while not reducing the maximum steering that can be distributed. Experimentally, one can verify the Gaussian steerablity via the minimum EPR variance product $E_{B'|A'}=(\cosh{2t}+1) /(2\cosh{2t})$ by homodyne detection. One finds $E_{B'|A'}<1$ for any $t>0$.

{\it Discussion.---} From Eq.~(\ref{GAtoB}), we find that the Gaussian steerability from Alice to Bob with displacement $d_B^{opt}$ can be distributed for any $t>0$; however, we need also check the separability of the states in {\it Step 2} and {\it Step 3} to assure that the transmitted ancilla stays separable from the rest at all stages.
\begin{figure}[tb]
\begin{center}
\includegraphics[width=0.75\columnwidth]{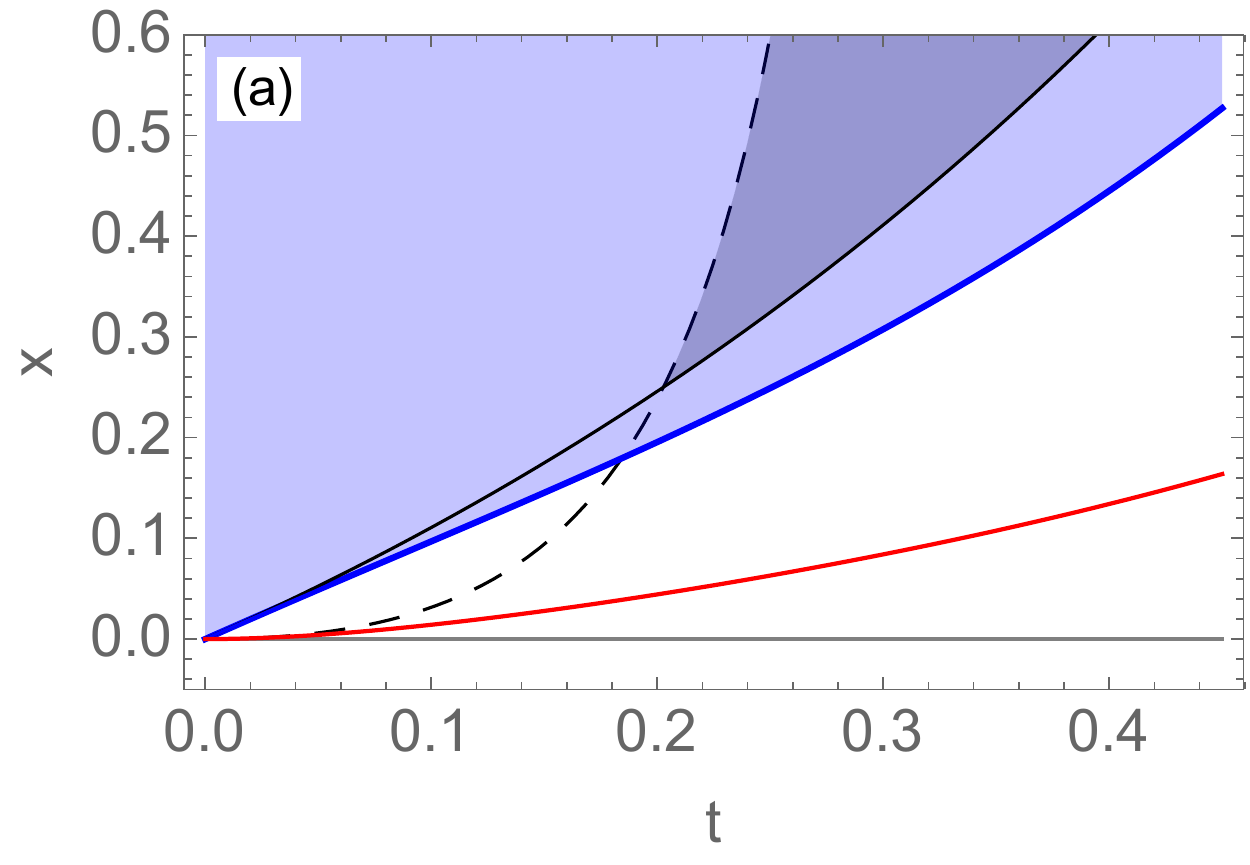}
\includegraphics[width=0.78\columnwidth]{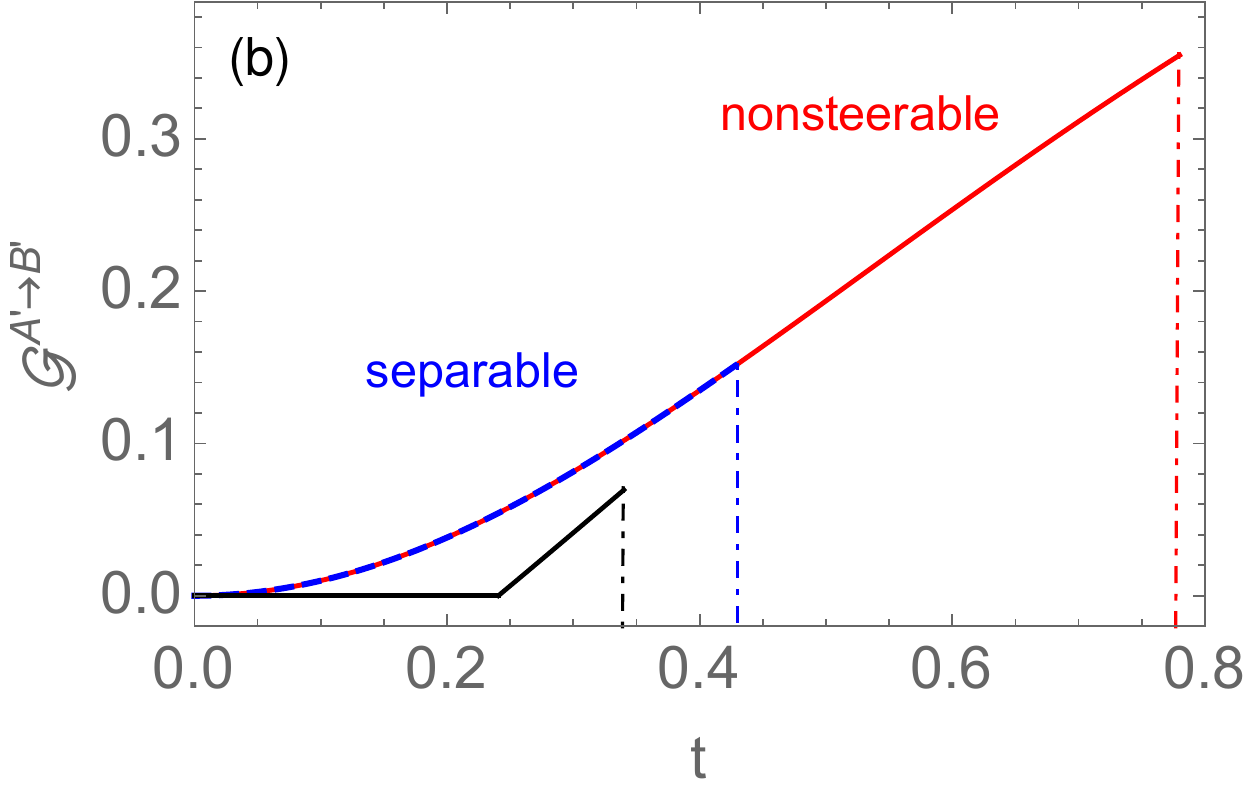}
\protect\caption{(a) The blue solid curve represents the threshold of $x_{C'-(A'B)}$, above which the state is separable with respect to $C'-(A'B)$ splitting. Since $\mathcal{G}^{A'\rightarrow B' }>0$ for any $t>0$ with optimal $d_B^{opt}$, the parameters within this area can be used to distribute Gaussian steering from mode $A'$ to mode $B'$ by transmitting a separable mode $C'$. With a non-optimal $d_B=2$, a much narrower range of parameters within the dark blue area between $x_{A\rightarrow B}$ (black dashed) and $x_{sep}$ (black solid) can be used to distribute Gaussian steering from Alice to Bob. The red solid curve shows a relaxed threshold for absence of steering instead of entanglement across $C'-(A'B)$ splitting in {\it Step 2}, above which one can use a nonsteerable mode $C'$ to distribute steering. (b) The distributed steerability $\mathcal{G}^{A\rightarrow B}$ for $x=0.5$ in the protocol with optimal
$d^{opt}_B$ and separable (blue dashed) or nonsteerable (red solid) ancilla, and in the case of non-optimal $d_B=2$ (black solid).
\label{AtoBsep}}
\end{center}
\end{figure}

After {\it Step 2}, the shared state transforms from a fully separable state to a two-mode biseparable state following the classification of Ref.~\cite{defineP}. Making use of the positive partial transpose (PPT) criterion~\cite{SimonPPT} one finds that the state with CM~(\ref{gamma2}) is entangled across $A'-(BC')$ splitting for any $x>0$ and $t>0$, but it is separable with respect to $C'-(A'B)$ splitting when $x \geq x_{C'-(A'B)}=2\cosh^2{2t}\sinh{t}/(\cosh{t}+\cosh{3t}+\sinh{t})$ (see blue solid curve in Fig. \ref{AtoBsep}(a)), and furthermore, it is separable with respect to $B-(A'C')$ splitting for any $x>0$ and $t>0$. Since steering is strictly stronger than entanglement, the CM~(\ref{gamma2}) also represents a state that is nonsteerable with respect to $C'-(A'B)$ splitting if $x \geq x_{C'-(A'B)}$ and $B-(A'C')$ splitting for any $x>0$ and $t>0$. However, the steerability $\mathcal{G}^{A'\rightarrow (BC')}>0$ for all $x>0$ and $t>0$, which is essential for the performance of the steering distribution from Alice to Bob in the final state. Without the help of the transmitted mode $C'$, the second beam splitter alone cannot create steering. In the blue area above the blue solid curve $x \geq x_{C'-(A'B)}$, the state after {\it Step 3} described by the CM~(\ref{gamma3}) remains separable with respect to $C''-(A'B')$ splitting. Therefore in this area, the ancilla mode is separable from the rest at all stages, nevertheless for the Gaussian steerablity of the final state one gets $\mathcal{G}^{A'\rightarrow B'}>0$ for any $t>0$, which means that Gaussian steering is successfully distributed from Alice to Bob. If we relax the condition that the ancilla is separable to that it is nonsteerable from the rest (i.e., it may be entangled), then the distribution task can be accomplished in an even larger region of parameters, as shown in the area above the red curve in Fig.~\ref{AtoBsep}(a).

Figure~\ref{AtoBsep}(b) shows the amount of one-way Gaussian steerability that can be distributed via sending a separable ancilla (blue dashed) and a nonsteerable ancilla (red solid), respectively. One can find that both of them distribute equal steerability from Alice to Bob, but work at different range of initial squeezing parameter $t$ for a fixed value $x=0.5$. By sending a nonsteerable ancilla, the initial squeezing level is requested to satisfy $0<t<0.78$ to guarantee that the transmitted mode $C'$ is nonsteerable from $(A'B)$ at all stages; while by sending a separable ancilla, the initial squeezing level is requested to satisfy a more stringent inequality $0<t<0.43$.

Comparing previous results with those for a non-optimal displacement, say~$d_B=2$, one finds that the distribution of steering via a separable ancilla can only work in the range of $x_{sep}=(e^{2t}-1)/2\leq x< x_{A\rightarrow B}=(1-e^{2t})^2/(4-2e^{2t})$ depicted by the dark blue area between the black solid and dashed curves in Fig.~\ref{AtoBsep}(a), which is much narrower than the area corresponding to the optimal $d^{opt}_B$. In addition, the distributed steerability (black solid in Fig.~\ref{AtoBsep}(b)) is also lower than that  given by the optimal protocol (blue dashed curve). For a fixed value of $x=0.5$, the protocol with $d_B=2$ can only work for squeezings obeying $0.241<t<0.346$, thus requiring a nontrivial threshold as opposed to the condition $t>0$ for optimal $d^{opt}_B$.

We have discussed distribution of Gaussian steering from Alice to Bob with a separable or nonsteerable ancilla. Can the distributed state simultaneously display also Gaussian steering from Bob to Alice in the setup given above? The answer is no. According to the CM (\ref{gamma3}), mode $B'$ and mode $C''$ are completely symmetric in the final state. Restricted by the monogamy relation of Gaussian steering with two observables $\hat{x}$ and $\hat{p}$ \cite{Reidmono}, neither of them can steer mode $A'$, so that only one-way Gaussian steering (under Gaussian measurements) from Alice to Bob is distributed using the above setup. If Bob wants to steer Alice, he can send the request to Could Server, and the Server can simply switch the initial quantum states and displacements for Alice and Bob.

\begin{figure}[tb]
\begin{center}
\includegraphics[width=1.\columnwidth]{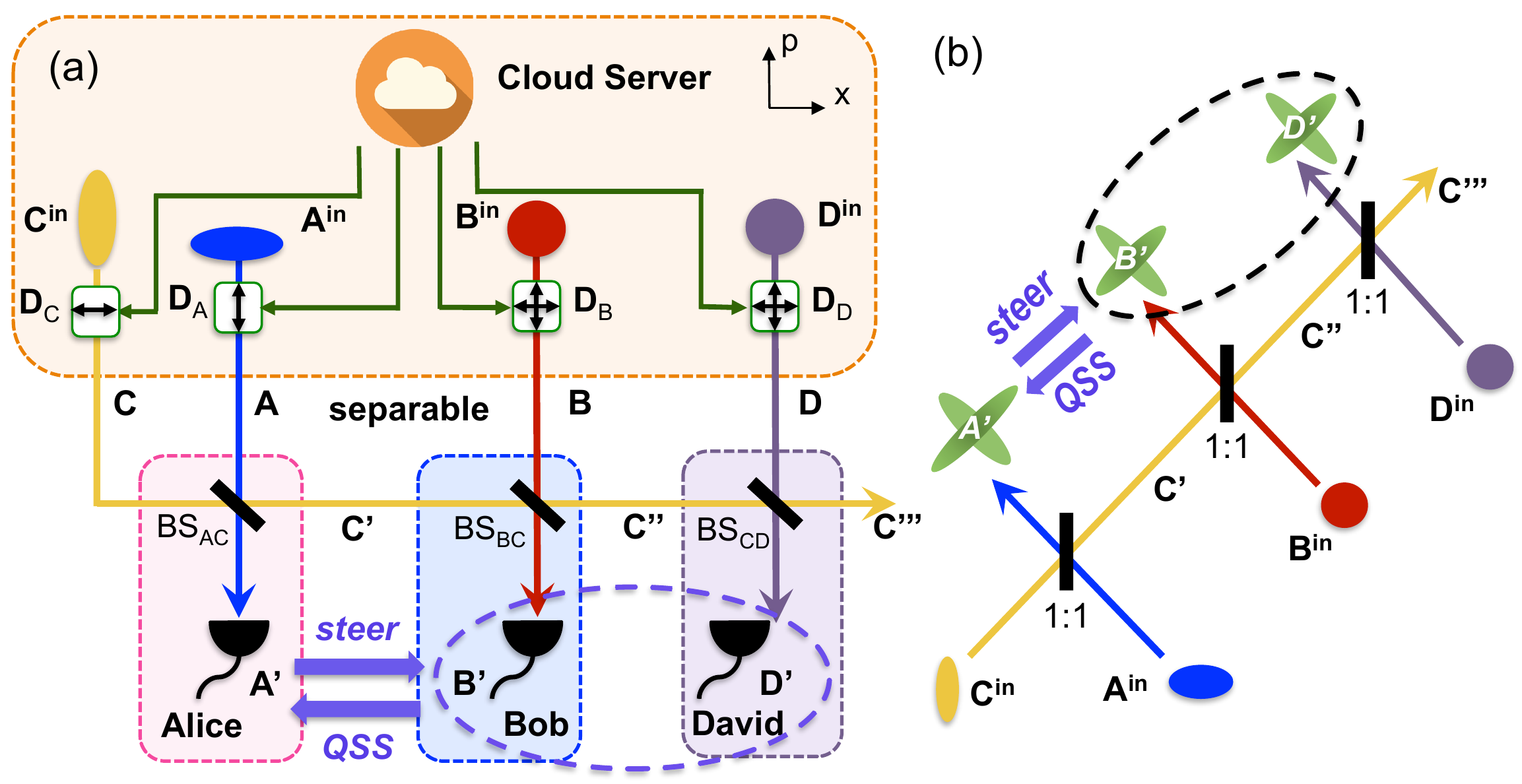}
\protect\caption{(a) Scheme for distribution of tripartite Gaussian steering via a separable state.
(b) Optical network with the same amount of steering as the optimal scheme with displacements. \label{system4}}
\end{center}
\end{figure}

{\it Multi-user distribution.---} The previous scheme can be extended to the multi-user case as shown in Fig.~\ref{system4}(a). Bob continues to send the separable mode $C''$ to David who mixes it with his mode $D$ on a balanced beam splitter $BS_{CD}$. It not only distributes tripartite steering from Alice to Bob and David, but also creates a collective steering in the opposite direction. 

To accomplish steering distribution in the direction $A'\rightarrow B'D'$, apart from the condition $x\geq x_{C'-(A'B)}$ that assures the mode $C'$ to be separable from modes $(A'B)$ after {\it Step 2}, we also need to find further constraints on $x$ and $t$ guaranteeing that the ancilla mode $C''$ is separable from subsystem $(A'B'D)$ when $x\geq x_{C''-(A'B'D)}$ after {\it Step 3}. Besides, we also need to suitably adjust $d_D$ in the displacement vectors $q_1=(0, -1, 0, d^{opt}_B, 0,-1, 0, d_D)^{\top}$ and $q_2 = (1, 0,d^{opt}_B, 0, -1, 0, d_D, 0)^{\top}$ to distribute steerability as large as possible. The CM of the resulting four-mode state is detailed in the Appendix~\cite{supp}. One can prove that with optimal displacement $d^{opt}_D=\sqrt{2}d^{opt}_B$, the highest distributed steerability reads: $\mathcal{G}^{A'\rightarrow B'D'}=\ln[4\cosh{2t}/(3+\cosh{2t})]$, $\mathcal{G}^{A'\rightarrow D'}=\ln[4\cosh{2t}/(1+3\cosh{2t})]$, $\mathcal{G}^{A'\rightarrow B'}=\ln[2\cosh{2t}/(1+\cosh{2t})]$,  and it can be achieved for any $t>0$. Since $x_{C'-(A'B)}>x_{C''-(A'B'D)}$ reported in ~\cite{supp},  in the blue area in Fig.~\ref{AtoBsep}(a) one can perfectly restore the steering of $A'\rightarrow B'D'$, $A'\rightarrow B'$, and $A'\rightarrow D'$ generated by the displacement-free optical network shown in Fig.~\ref{system4}(b)~\cite{supp}.

For the opposite direction  $B'D'\rightarrow A'$, keeping $d_B^{opt}$ we can distribute the maximum steerability $\mathcal{G}^{B'D'\rightarrow A'}= \ln[(1+3\cosh{2t})/(3+\cosh{2t})]$ with $d_D^{opt}=(2+2\coth{t}+\tanh{t}-\tanh{2t})/\sqrt{2}$, which is recovering the steering created in Fig.~\ref{system4}(b). Note that the protocol works only when $t\geq 0.943$ and $x\geq\mathrm{max}\{x_{C'-(A'B)},\ x_{C''-(A'B'D)}\}=x_{C''-(A'B'D)}$~\cite{supp} to assure the state to be separable with respect to $C'-(A'B)$ splitting, as well as $C''-(A'B'D)$ splitting. For smaller $t$, we need to optimize $d_B$ and $d_D$ simultaneously. We find that when $0.28\leq t<0.943$, the distributed steering $\mathcal{G}^{B'D'\rightarrow A}$ can be still maximized by choosing some appropriate displacements $d_B$ and $d_D$ through numerical search, while when $t<0.28$, it's impossible to achieve the same amount of steerability as in the scheme in Fig.~\ref{system4}(b)~\cite{supp}. In this direction, $\mathcal{G}^{B'\rightarrow A'}=\mathcal{G}^{D'\rightarrow A'} =0$, which means that neither Bob nor David can individually steer Alice, but they can do that only if they collaborate. 

This makes the state a perfect resource for the 1sDI QSS protocol, where Alice doesn't trust Bob and David's devices. Assume Alice acts as the dealer who sends a secret encoded in her state, while Bob and David are players aiming at decoding the message together. To provide security against eavesdropping, a guaranteed secret key rate for the QSS protocol is given by $K\geq \ln[2/(eE_{A'|B'D'})]=\mathcal{G}^{B'D'\rightarrow A'}-\ln(e/2)$~ \cite{YuQSS,GiannisQSS}. A state whose correlations result in $K > 0$ can be regarded a useful resource. Referring to the studied scheme, the condition translates into
$\cosh{2t} > (3e-2)/(6-e)$, which means that a squeezing level of $5.4$dB ($t > 0.62$), is required to ensure a nonzero key rate. This is well within the current experimental feasibility, since up to $10$dB of squeezing has been demonstrated~\cite{8db,10db}.

{\it Summary.---} We proposed a protocol for distribution of Gaussian EPR steering between distant parties with a separable ancilla. The rich steering properties, such as one-way, one-to-multimode and collective Gaussian steering, can be distributed via local operations on parts of an initial
fully separable state and communication of a separable part of the state. We derived analytical thresholds on input squeezing and displacement strength needed for accomplishment of various Gaussian steering distribution protocols. Moreover, we have shown that with optimal displacements the largest steerability that can be distributed coincides with the steering generated by the same optical network but without displacements on input states. Our work shows that a key quantum feature such as EPR steering, which is strictly stronger than entanglement, can be faithfully distributed across multi-user networks with minimal resources. The proposed protocols can be implemented by performing suitable local correlated displacements on the input states of a linear optical networks which are usually used to generate multipartite CV entangled states and hence they are feasible with current technology.

\textbf{Acknowledgments} Y.X., X.S. and Q.H. acknowledge the support of the Ministry of Science and Technology of China (Grants No. 2016YFA0301302 and No. 2016YFA0301402) and the National Natural Science Foundation of China (Grants No. 11622428, No. 11522433, No. 61475006, No. 61675007, and No. 61475092,). X.S. thanks the program of Youth Sanjin Scholar and the Fund for Shanxi ``1331 Project'' Key Subjects Construction. G.A. acknowledges funding from the European Research Council under Starting Grant No. 637352 (GQCOP).
\appendix*
\setcounter{equation}{0}
\subsection*{Appendix A: Details of schemes without displacement}

In the main text, we show that the largest steerability that can be distributed recovers that of the multimode states created by the same optical network without the correlated displacements, shown in Fig.~2(b) and Fig.~4(b). This means the displacements only make the transmitted ancilla mode separable from the rest without reducing the amount of steering. Here, we give out the details of the covariance matrixes of the three-mode and four-mode states generated in Fig.~2(b) and Fig.~4(b) as follows.

As shown in Fig.~2(b), the input states $A^{\rm{in}}$ and $C^{\rm{in}}$ are in momentum- and position-squeezed vacuum states, respectively, while mode $B^{\rm{in}}$ is in a vacuum state, which are same with the initial states prepared by Cloud Server in Fig.~2(a). Through two balanced beam splitters, the covariance matrix of output state becomes
\begin{equation}
\gamma_{A'B'C'}=\left(
\begin{array}{ccc}
\cosh{(2t)}\mathds{1}  &\frac{\sqrt{2}\sinh{(2t)}}{2}\sigma_z &\frac{-\sqrt{2}\sinh{(2t)}}{2}\sigma_z\\
\frac{\sqrt{2}\sinh{(2t)}}{2}\sigma_z &\cosh^2{t}\mathds{1} &-\sinh^2{t}\mathds{1}\\
\frac{-\sqrt{2}\sinh{(2t)}}{2}\sigma_z & -\sinh^2{t}\mathds{1} & \cosh^2{t}\mathds{1}
\end{array}
\right). \label{gamma2b}
\end{equation}
By quantifying the amount of the Gaussian steering from mode $A'$ to mode $B'$ via Eq.~(1) in the main text, we can get same steerability as shown in Eq.~(6), i.e., the distributed Gaussian steering $\mathcal{G}^{A'\rightarrow B'}$ by transmitting a separable ancilla mode with optimal displacements.

Similarly, in the Multi-user case, we add an additional vacuum state $D^{\rm{in}}$ and a balanced beam splitter, as shown in Fig.~4(b) in the main text. The covariance matrix of the output four-mode state becomes 
\begin{equation}
\gamma_{A'B'C'D'}=
\left(
\begin{array}{cccc}
\cosh{(2t)}\mathds{1}  &\frac{\sqrt{2}\sinh{(2t)}}{2}\sigma_z &\frac{-\sinh{(2t)}}{2}\sigma_z&\frac{-\sinh{(2t)}}{2}\sigma_z\\
\frac{\sqrt{2}\sinh{(2t)}}{2}\sigma_z &\cosh^2{t}\mathds{1} &\frac{-\sinh^2{t}}{\sqrt{2}}\mathds{1}&\frac{-\sinh^2{t}}{\sqrt{2}}\mathds{1}\\
\frac{-\sinh{(2t)}}{2}\sigma_z & \frac{-\sinh^2{t}}{\sqrt{2}}\mathds{1} & \frac{3+\cosh{(2t)}}{4}\mathds{1}&\frac{\sinh^2{t}}{2}\mathds{1}\\
\frac{-\sinh{(2t)}}{2}\sigma_z & \frac{-\sinh^2{t}}{\sqrt{2}}\mathds{1} & \frac{\sinh^2{t}}{2}\mathds{1}&\frac{3+\cosh{(2t)}}{4}\mathds{1}
\end{array}
\right). \label{gamma4b}
\end{equation}
Using Eq.~(1) in the main text, we can quantify the Gaussian steerability $\mathcal{G}^{A'\rightarrow B'D'}$, $\mathcal{G}^{A'\rightarrow B'}$, $\mathcal{G}^{A'\rightarrow D'}$ in the direction $A'\rightarrow B'D'$, and the Gaussian steerability $\mathcal{G}^{B'D'\rightarrow A'}$ in the opposite direction. One can find that the amount of steerability in this optical network characterized by CM (\ref{gamma4b}) are same with the present distribution protocol via separable ancilla mode with optimal displacements shown in Fig.~4(a), as shown in the next section.

\subsection*{Appendix B: Details of Multi-user distribution}

The present protocol can be extended to quantum network with multi-user as shown in Fig.~4(a) in main text. After \textit{Step 3}, Bob continues to send the separable ancilla mode to next user David who mixes it with his mode on a balanced beam splitter. The three users can successfully share EPR steering with desired properties, not only tripartite steering from Alice to Bob and David, but also a collective steering in the opposite direction from Bob and David to Alice. The details of the covariance matrix of the four-mode states and the optimal displacement on each mode are given as follows.

The CM of the state after the third step becomes
\begin{equation}
\gamma_{A'B'C''D}=\left(
\begin{array}{cc}
\gamma_{A'B'C''} & \varepsilon  \\
\varepsilon^{\top}  & \gamma_{D}
\end{array}
\right), \label{gamma4}
\end{equation}
where $\gamma_{A'B'C''}$ still maintains the form of Eq.~(6) in the main text, and $\gamma_{D}=(1+d_D^2x)\mathds{1}$, $\varepsilon=(xd_D\sigma_z,~(d_B-1)d_Dx\mathds{1}/\sqrt{2},~(d_B+1)d_Dx\mathds{1}/\sqrt{2})^{\top}$.

Then David mixes the ancilla mode $C''$ sent by Bob with his mode $D$ on the balanced beam splitter $BS_{CD}$, which yields the final CM $\gamma_{A'B'C'''D'}=U_{CD}\gamma_{A'B'C''D}U_{CD}^{\top}$ in the form
\begin{equation}
\gamma_{A'B'C'''D'}=\left(
\begin{array}{cccc}
m\mathds{1}  &l\sigma_z & f\sigma_z  &g\sigma_z\\
l\sigma_z & s\mathds{1} & h\mathds{1} &j\mathds{1}\\
 f\sigma_z &h\mathds{1}&k\mathds{1}&v\mathds{1}\\
g\sigma_z & j\mathds{1} &v\mathds{1}& w\mathds{1}
\end{array}
\right), \label{gamma5}
\end{equation}
with
\begin{eqnarray}
&&l=\frac{d_Bx+n}{\sqrt{2}}, \quad s=\frac{1+m+d_Bx(d_B-2)}{2},\nonumber\\
&&f=\frac{(d_B+\sqrt{2}d_D)x-n}{2}, \quad g=\frac{(d_B-\sqrt{2}d_D)x-n}{2},\nonumber\\
&&h=\frac{(\sqrt{2}d_B^2+2d_Bd_D-2d_D)x-\sqrt{2}(m-1)}{4},\nonumber\\
&&j=\frac{(\sqrt{2}d_B^2-2d_Bd_D+2d_D)x-\sqrt{2}(m-1)}{4},\nonumber\\
&&k=\frac{3+m-x+(1+d_B+\sqrt{2}d_D)^2x}{4},\nonumber\\
&&v=\frac{m-1+(d_B^2+2d_B-2d_D^2)x}{4},\nonumber\\
&&w=\frac{3+m-x+(1+d_B-\sqrt{2}d_D)^2x}{4}.
\end{eqnarray}
The free parameters $d_B,\ d_D$ and $x$ need to be suitably adjusted for distributing the EPR steering with desired properties.
\begin{figure}
\begin{center}
\includegraphics[width=.75\columnwidth]{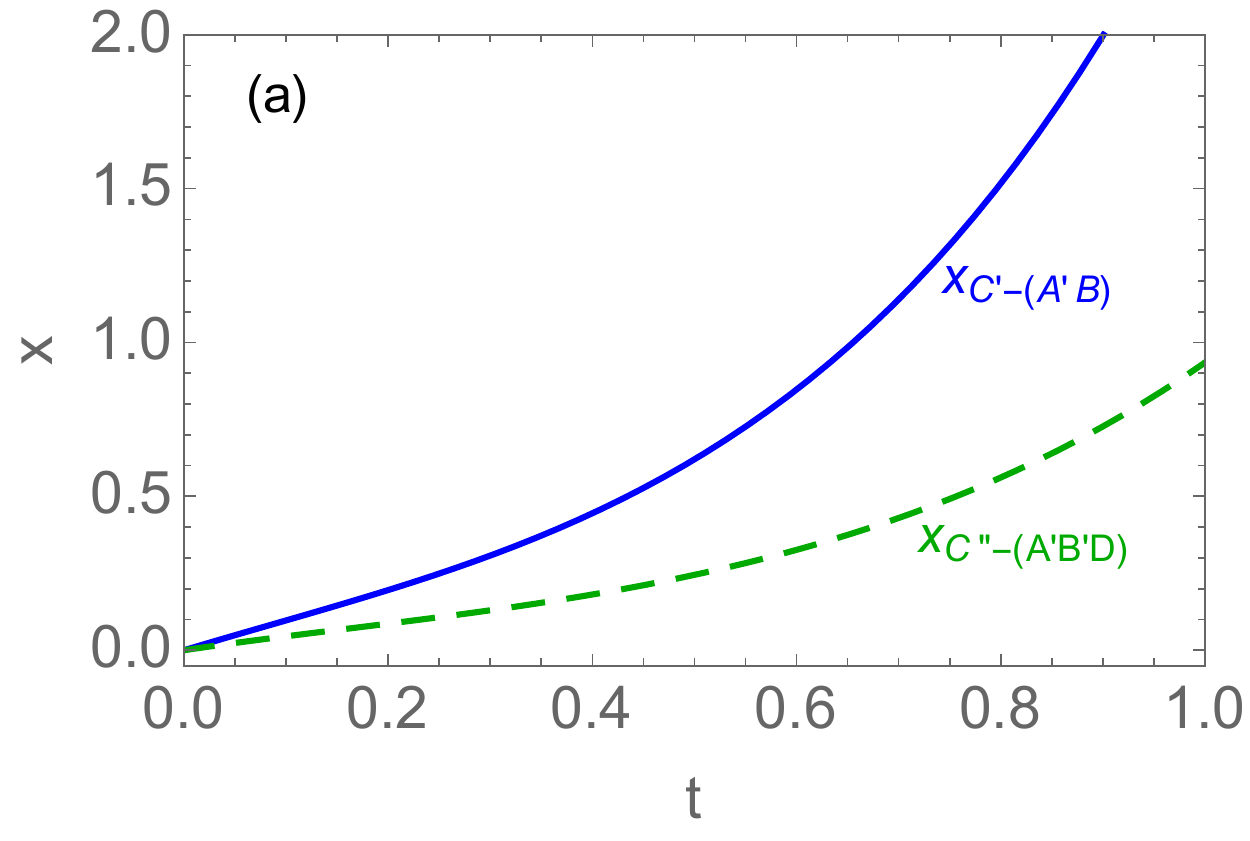}
\includegraphics[width=.75\columnwidth]{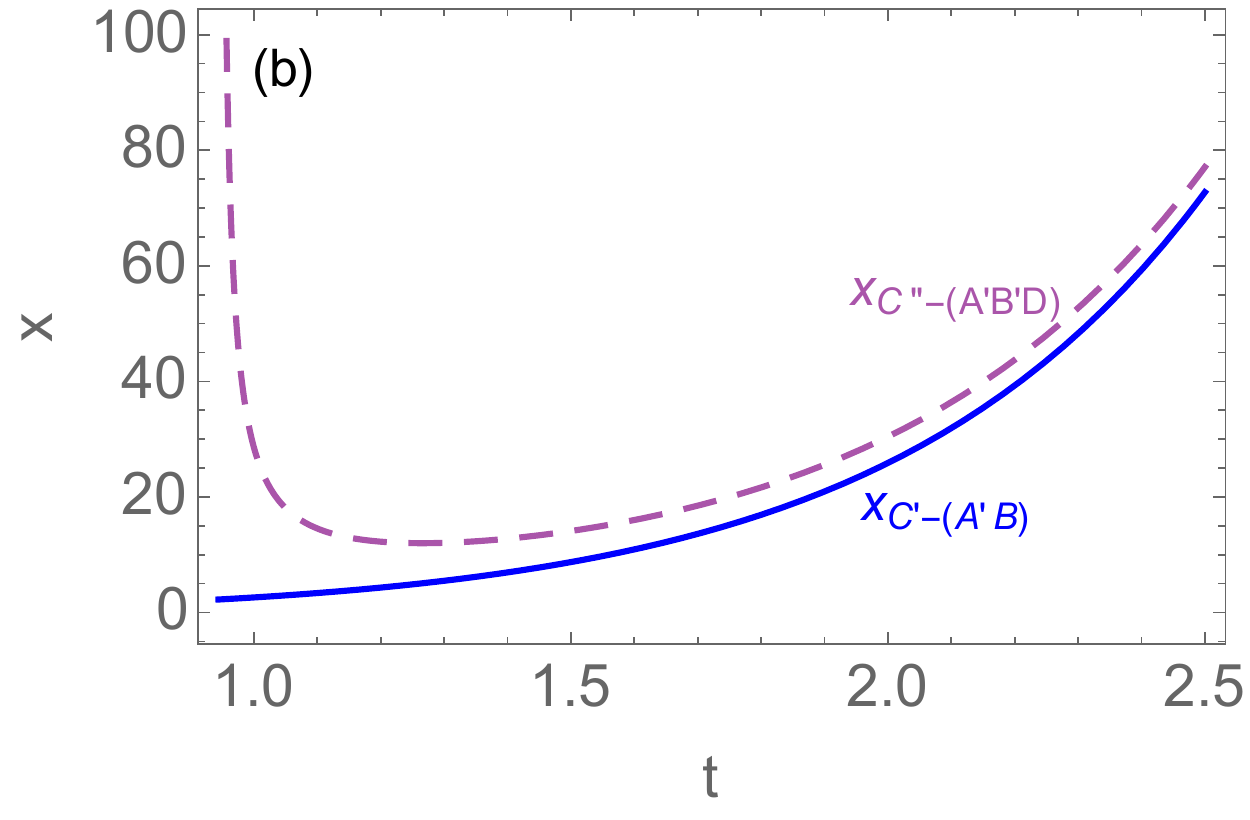}
\protect\caption{In the direction of (a) $A'\rightarrow B'D'$ and (b) $B'D'\rightarrow A'$, the range of $x$ as a function of $t$ to satisfy the separability requirements for $C'-(A'B)$ splitting and for $C''-(A'B'D)$ splitting.\label{x}}
\end{center}
\end{figure}

Based on CM~(\ref{gamma5}), the maximum of Gaussian steering can be distributed from Alice to Bob and David is given by
\begin{eqnarray}\label{GAtoBD}
\mathcal{G}^{A'\rightarrow B'D'}&=&\ln[4\cosh{2t}/(3+\cosh{2t})],\nonumber\\
\mathcal{G}^{A'\rightarrow D'}&=&\ln[4\cosh{2t}/(1+3\cosh{2t})], \nonumber\\
\mathcal{G}^{A'\rightarrow B'}&=&\ln[2\cosh{2t}/(1+\cosh{2t})],
\end{eqnarray}
with optimal displacements $d_B^{opt}=\tanh(2t)+1$ and $d_D^{opt}=\sqrt{2}[\tanh(2t)+1]=\sqrt{2}d^{opt}_B$. This is same to the amount of steering from $A'$ to $B'$ and $D'$ generated by the displacement-free optical network given in Fig. 4(b) in the main text.

To satisfy the conditions where the ancilla mode $C'$ is separable from the subsystem $(A'B)$ after {\it Step 2}, and the ancilla mode $C''$ is separable from the subsystem $(A'B'D)$ after {\it Step 3}, we need choose $x\geq\mathrm{max}\{x_{C'-(A'B)},\ x_{C''-(A'B'D)}\}$. From Fig.~\ref{x}(a), we can find that for any $t>0$, $x_{C'-(A'B)}>x_{C''-(A'B'D)}$, where $x_{C'-(A'B)}=2\cosh^2{2t}\sinh{t}/(\cosh{t}+\cosh{3t}+\sinh{t})$ and $x_{C''-(A'B'D)}=2\cosh^2{2t}\sinh{t}/[2(\cosh{t}+\cosh{3t}+\sinh{t})+\sinh{3t}]$, so that in the blue shadow area of Fig.~3 (a) in the main text one can perfectly restore the steering of $A'\rightarrow B'D'$, $A'\rightarrow B'$, and $A'\rightarrow D'$ produced by the optical network given in Fig.~4 (b) in the main text without making displacements on input states.

But for the other direction ($\mathcal{G}^{B'D'\rightarrow A'}$), the discussion is more complex. When $t<0.28$, we can distribute a nonzero collective steering via separable ancilla mode but it's impossible to fully recover the amount of steerability given in Fig.~4 (b) in the main text. When $0.28\leq t<0.943$, the distributed steering $\mathcal{G}^{B'D'\rightarrow A}$ can be maximized by choosing some appropriate displacements $d_B, d_D$ and $x$ through numerical search. The maximal steerability that can be distributed is
\begin{equation}\label{GAtoBD}
\mathcal{G}^{B'D'\rightarrow A'}=\ln[(1+3\cosh{2t})/(3+\cosh{2t})],
\end{equation}
which is recovering the steering created in Fig.~4 (b) in the main text. When $t\geq 0.943$, the design can be analytically optimized with $d_B^{opt}=\tanh(2t)+1$ and $d_D^{opt}=(2+2\coth{t}+\tanh{t}-\tanh(2t))/\sqrt{2}$ in the range of $x\geq\mathrm{max}\{x_{C'-(A'B)},\ x_{C''-(A'B'D)}\}$, which assures the ancilla modes $C'$ and $C''$ are separable from the rest modes. From Fig.~\ref{x}(b), we can find that for any $t\geq0.943$, $x_{C''-(A'B'D)}>x_{C'-(A'B)}$, where $x_{C'-(A'B)}=2\cosh^2{2t}\sinh{t}/(\cosh{t}+\cosh{3t}+\sinh{t})$ and $x_{C''-(A'B'D)}=2\sinh^2{4t}/(2\sinh{2t}-12\cosh{2t}-4\sinh{4t}-7\cosh{4t}+2\sinh{6t}-13)$.


\begin{thebibliography}{99}
\bibitem{noteGauss}Note that throughout the paper, the (non)steerability is always intended under the experimentally motivated restriction of Gaussian measurements; for steerability of Gaussian states by non-Gaussian measurements, see e.g.~\cite{Yunon}.
\end{thebibliography}
\end{document}